\documentclass[
 reprint, 
 amsmath,amssymb,
 aps,
 prl,
]{revtex4-2}
\usepackage{verbatim}
\usepackage{soul}
\usepackage{braket}
\usepackage{graphicx}
\usepackage{dcolumn}
\usepackage{bm}
\usepackage{xspace}
\usepackage{xcolor}
\newcommand{\am}{Auger-Meitner\xspace}
\newcommand{\pa}{phonon-assisted\xspace}
\newcommand{\amu}{cm\(^6\)s\(^{-1}\)\xspace}
\newcommand{\oee}{\(10^{18}\) cm\(^{-3}\)\xspace}
\newcommand{\cmc}{cm\(^{-3}\)\xspace}

\newcommand{\beginsupplement}{
            \setcounter{table}{0}
            \renewcommand{\thetable}{S\arabic{table}}
            \setcounter{figure}{0}
            \renewcommand{\thefigure}{S\arabic{figure}}
            \setcounter{equation}{0}
            \renewcommand{\theequation}{S\arabic{equation}}
}

\begin{document}
\title{Phonon-Assisted \am Recombination in Silicon from First Principles}
\author{Kyle Bushick}
\author{Emmanouil Kioupakis}
 \email{kioup@umich.edu}
\affiliation{
Department of Materials Science and Engineering, University of Michigan, Ann Arbor, Michigan 48109, USA
}

\date{July 10, 2023}

\begin{abstract}
We present a consistent first-principles methodology to study both direct and phonon-assisted Auger-Meitner recombination (AMR) in indirect-gap semiconductors that we apply to investigate the microscopic origin of AMR processes in silicon. Our results are in excellent agreement with experimental measurements and show that phonon-assisted contributions dominate the recombination rate in both \textit{n}-type and \textit{p}-type silicon, demonstrating the critical role of phonons in enabling AMR. We also decompose the overall rates into contributions from specific phonons and electronic valleys to further elucidate the microscopic origins of AMR. Our results highlight potential pathways to modify the AMR rate in silicon via strain engineering.
\end{abstract}

\maketitle

\begin{figure}[ht]
\includegraphics{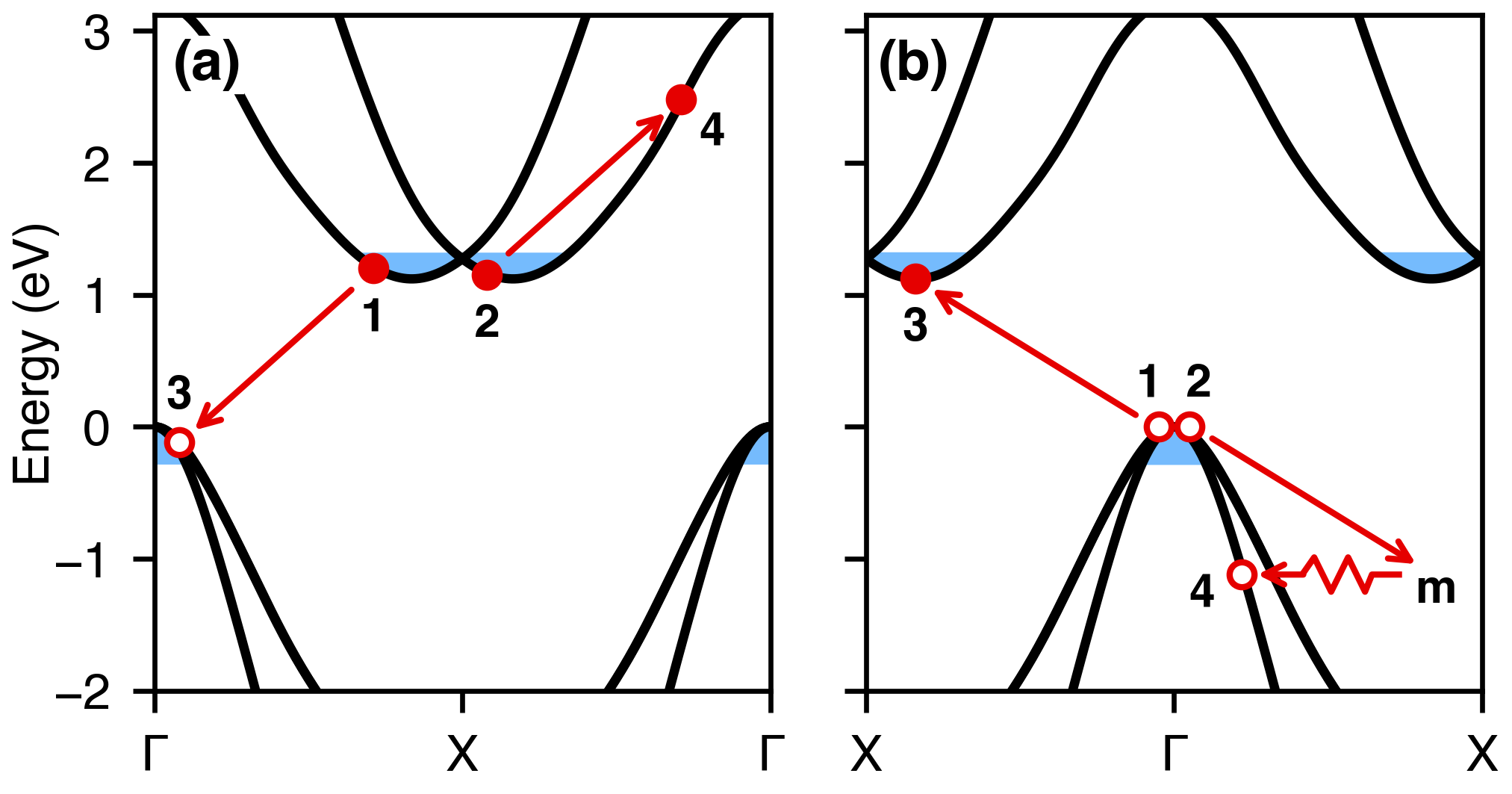}
\caption{Quasiparticle band structure of silicon depicting Auger-Meitner recombination processes. (a) Schematic of the direct \(eeh\) \am process involving two electrons (solid circles) near the conduction band minimum and a hole (open circle) near the valence band maximum. The blue areas indicate the range of initial carrier states that are included in the calculations. (b) Schematic of the \pa \(hhe\) \am process, where the inclusion of a phonon conserves the overall momentum and allows this process to take place.}
\label{fig:bands}
\end{figure}

\am recombination (AMR), also referred to as Auger recombination in the literature, is an intrinsic non-radiative carrier recombination process in semiconductors that is named after Lise Meitner and Pierre Auger. AMR of free carriers in bulk materials parallels the atomic effect, whereby a core hole is filled by an electron, transferring the excess energy to a second, ejected electron. This atomic emission process was first discovered by Meitner in 1922, and independently by Auger in 1923, bearing both their names in recognition of their contributions.\cite{Matsakis2019} In the semiconductor AMR process, an electron and a hole recombine across the band gap, transferring their energy via the Coulomb interaction to another electron (\(eeh\) process) or hole (\(hhe\) process) and exciting it to a high energy state. The AMR mechanism can occur in a direct fashion if the carriers can satisfy both energy and momentum conservation (Fig. \ref{fig:bands}a). Alternatively, \pa AMR occurs when momentum conservation is satisfied through the absorption or emission of a phonon (Fig. \ref{fig:bands}b), much like phonon-assisted optical absorption in indirect-gap materials. The additional momentum provided by the phonon increases the number of final electronic states accessible to AMR, making the \pa process dominant in cases where direct AMR is weak or not possible. The AMR rate for nondegenerate free carriers is proportional to the third power of \(n\) (\(R=\frac{dN}{dt}=CVn^3\)), where \(n = \frac{N}{V}\) is the number of free carriers \(N\) per volume \(V\), and \(C\) is the AMR coefficient. AMR is of broad interest as it has been shown to limit the maximum efficiency of solar cells,\cite{Tiedje1984,Kerr2003} LEDs,\cite{Shen2007} bipolar transistors,\cite{Tyagi1983} lasers,\cite{Takeshima1985,Singh1995} the ideality factor in diodes,\cite{Leilaeioun2016} and becomes the dominant recombination pathway at high carrier concentrations.

Despite its scientific and engineering importance, direct and \pa AMR remains poorly understood in materials {\textendash} including silicon. While there have been a number of experimental studies and parameterizations of the overall AMR rate in silicon,\cite{Altermatt1997,Kerr2002,Richter2012a,Black2022} important fundamental questions about the process remain unanswered. For one, empirical models rely on challenging measurements and empirical function fits, preventing them from probing the underlying atomistic mechanisms. For example, empirical models are not capable of separating the effects of the different varieties of AMR (e.g., direct or \pa) or evaluating different microscopic contributions (e.g., from different electron valleys or specific phonon modes). These aspects limit the utility of such models in gaining a fundamental understanding of AMR in silicon.

Considering the constraints of empirical fits to experimental data, theoretical calculations offer an alternative route to probe the underlying mechanisms of AMR. However, past work in this area has been limited by a variety of shortcomings. Early efforts from Huldt investigated the direct AMR process but used a number of approximations to both the band structure and matrix elements, leading to underestimates of the AMR rate.\cite{Huldt1971} In follow-up work, Lochmann and Haug also considered phonon-assisted processes, but used a semi-empirical band structure, limited the electron-phonon scattering mechanisms and recombination pathways, estimated the matrix elements with \(k\cdot{p}\), and omitted Umklapp scattering.\cite{Lochmann1978,Haug1978,Lochmann1980} These simplifications exclude key components of the physical process and can lead to inaccurate results {\textendash} a finding discussed in greater detail by Laks et al.\cite{Laks1988,Laks1990}. In their subsequent work on direct AMR, Laks et al. used an empirical pseudopotential and discarded the previous simplifying assumptions, concluding that the direct AMR process is dominant for \(eeh\) AMR, though it is insufficient to describe the \(hhe\) AMR rate. Govoni et al. later performed the only first-principles calculation for direct AMR in silicon, finding loose agreement with experimental results for \(eeh\) AMR and attributing the difference to \pa AMR.\cite{Govoni2011} To date, there has not been a full band structure calculation that includes both direct and \pa processes, a limit imposed by the computational complexity of the problem. This shortcoming has precluded definitive conclusions regarding the microscopic mechanisms of AMR in silicon that are resolved by our work. 

In this Letter, we implement a consistent first-principles computational methodology to investigate both direct and \pa AMR in indirect-gap materials at the same level of theory and apply it to address the long-standing challenge of gaining a complete and accurate mechanistic understanding of AMR in silicon. Our results show excellent agreement with experiment, including both the carrier and temperature dependence of the AMR coefficient in this technologically important material. We demonstrate that the \pa mechanism is dominant, not only for \textit{p}-type silicon, but also for \textit{n}-type silicon, for which direct AMR is also possible. Furthermore, we probe the electron-valley dependence for the direct and \pa \(eeh\) processes, and we propose epitaxial strain as a new route for modulating the AMR rate in silicon. Our methodology for accurately assessing both direct and \pa AMR processes can also be readily applied to other material systems and advance our atomistic understanding of AMR in electronic and optoelectronic semiconductor devices.

The direct AMR coefficient can be evaluated using first-order perturbation theory and Fermi's golden rule, while the \pa AMR process requires second-order perturbation theory. We direct the reader to Ref. \citenum{Kioupakis2015} for these formulations, which we also summarize in the Supplemental Material (including Ref. \citenum{Cappellini1993}).\cite{supplemental} The summation in Eq. S2 is computationally intractable if we account for the carrier distribution around the band extrema, as depicted in Fig. \ref{fig:bands}b, as the additional momentum of the phonon relaxes the momentum conservation constraint. Thus, to enable the calculation of the \pa AMR rate, we assume that all low-energy carriers have the energy, momentum, and wave function of the states at the corresponding band extrema. For indirect gap materials, however, care must be taken to deal with the band and valley degeneracies properly. We can rewrite Eq. S2 as
\begin{align}
\begin{split}
     R_{pa,eeh} &= 2\frac{2\pi n_e^2 n_h}{8\hbar N_\mathbb{C}^2 N_\mathbb{V}}\sum_{\mathbf{1234};\nu\mathbf{q}}(n_{\nu\mathbf{q}}+\frac{1}{2}\pm\frac{1}{2})\\&\times|\tilde{M}_{\mathbf{1234};\nu\mathbf{q}}|^2\delta(\epsilon_\mathbf{1}+\epsilon_\mathbf{2}-\epsilon_\mathbf{3}-\epsilon_\mathbf{4}\mp\hbar\omega_{\nu\mathbf{q}}),
\end{split}
\label{eq:eeh}
\end{align}
for the \(eeh\) process, where \(n_e\) and \(n_h\) are the electron and hole carrier concentrations, \(N_\mathbb{C}\) is the conduction band minimum total degeneracy, defined as \(N_\mathbb{C}\equiv N^\mathbb{C}_{valley}\times N^\mathbb{C}_{band}\), and \(N_\mathbb{V}\) is the valence band maximum total degeneracy, defined as \(N_\mathbb{V}\equiv N^\mathbb{V}_{valley}\times N^\mathbb{V}_{band}\). The \(R_{pa,hhe}\) term simply swaps the powers of \(N_\mathbb{C}\) and \(n_e\) with \(N_\mathbb{V}\) and \(n_h\) in the prefactor. For silicon, \(N^\mathbb{C}_{valley}=6\), \(N^\mathbb{C}_{band}=1\), \(N^\mathbb{V}_{valley}=1\), and \(N^\mathbb{V}_{band}=3\). We note that \(N^\mathbb{V}_{band}=3\) because we do not consider spin-orbit coupling in our calculations. Previous work on AMR in InAs found that including spin-orbit coupling did not appreciably alter the AMR rate unless the splitting was large enough to form resonant states with the band gap.\cite{Shen2019} Given that the spin-orbit splitting in silicon is significantly smaller than the band gap, we can safely omit it from our calculations in order to reduce the computational cost without suffering any severe loss of accuracy. 

We perform our calculations using a range of first-principles tools. We utilize the open-source \texttt{Quantum ESPRESSO} package to obtain density functional theory (DFT) wave functions and eigenvalues for the DFT relaxed structure (\(a=5.379\) \AA) as well as density functional perturbation theory (DFPT) calculations to obtain the electron-phonon coupling matrix elements.\cite{Giannozzi2009,Giannozzi2017} We also employ the \texttt{BerkeleyGW} code to calculate the G\textsubscript{0}W\textsubscript{0} (GW) quasiparticle corrections to DFT eigenenergies and band curvature.\cite{Hybertsen1986a,Deslippe2012} Since AMR must be calculated on much finer grids than are required for the band structure, we employ the maximally localized Wannier function method and the \texttt{wannier90} package to interpolate our GW eigenvalues onto arbitrarily fine grids.\cite{Pizzi2020} It is known that even these higher levels of theory do not recover the exact effective masses,\cite{Ponce2018} so we also conduct sensitivity tests of the effective mass on the AMR coefficient, finding that even 10\% changes to the effective mass do not appreciably affect the results. While the GW band gap is sensitive to temperature, we show that the AMR coefficient is not strongly dependent on the band gap and therefore do not include these temperature effects in our calculations. Convergence testing with respect to the Brillouin zone (BZ) sampling also demonstrates that our AMR rates are converged within 5\% for the direct \(eeh\) process and both \pa processes and 11\% for the direct \(hhe\) process (which we show to be negligible). Details on these tests, as well as the other computational parameters and their convergence can be found in the Supplemental Material,\cite{supplemental} which includes Refs. \citenum{Okada1984,Deslippe2013,Bludau1974,Brown2016}.

\begin{table*}[t]
\centering
\begin{tabular}{ccccccc}
Source & \(C_{eeh,dir}\) & \(C_{eeh,pa}\) & \(C_{eeh,tot}\) & \(C_{hhe,dir}\) & \(C_{hhe,pa}\) & \(C_{hhe,tot}\) \\ \hline
This Work                                             & 0.86           & 2.33 & 3.19  & 0.000089           & 2.0 & 2.0  \\
Govoni, Marri, and Ossicini (theory)\cite{Govoni2011} & \({\sim}1.07\) & \textendash & \textendash & \({\sim}0.000049\) & \textendash & \textendash \\
Laks, Neumark, and Pantelides (theory)\cite{Laks1990} & \({\sim}3.9\)  & \textendash & \textendash & \({\sim}0.015\)    & \textendash & \textendash \\
Dziewior and Schmid (experiment)\cite{Dziewior1997}   & \textendash    & \textendash & 2.8  & \textendash               & \textendash & 0.99 \\
H\"{a}cker and Hangleiter (experiment)\cite{Hacker1994}   & \textendash    & \textendash & 4.35 & \textendash               & \textendash & 2.02  
\end{tabular}
\caption{Comparison of calculated and measured AMR coefficients. All data are shown in units of \(\times10^{-31}\) \amu at a temperature of 300 K and carrier concentrations in the range of \oee.}
\label{tab:amr}
\end{table*}

The summary of our combined direct and \pa AMR coefficients in silicon are shown in Table \ref{tab:amr}. Our direct calculations agree well with the first-principles results from Govoni, Marri, and Ossicini, which are the most accurate calculations of direct AMR in silicon to date.\cite{Govoni2011} We also find superb agreement with the experimentally measured values from both Dziewior and Schmid and H\"{a}cker and Hangleiter.\cite{Dziewior1997,Hacker1994} Because of the relatively close agreement between past direct \(eeh\) calculations and experiment, direct AMR has typically been considered sufficient to explain the measurements in \textit{n}-type silicon.\cite{Laks1988,Laks1990,Govoni2011} Contrasting this, our findings not only confirm that \pa AMR is the dominant mechanism for the \(hhe\) process (\({>}99.9\%\) of the total), but also for \(eeh\) AMR (\({\sim}73\%\)), demonstrating the importance of phonons to both \(eeh\) and \(hhe\) AMR processes in silicon.

\begin{figure}[ht]
\includegraphics{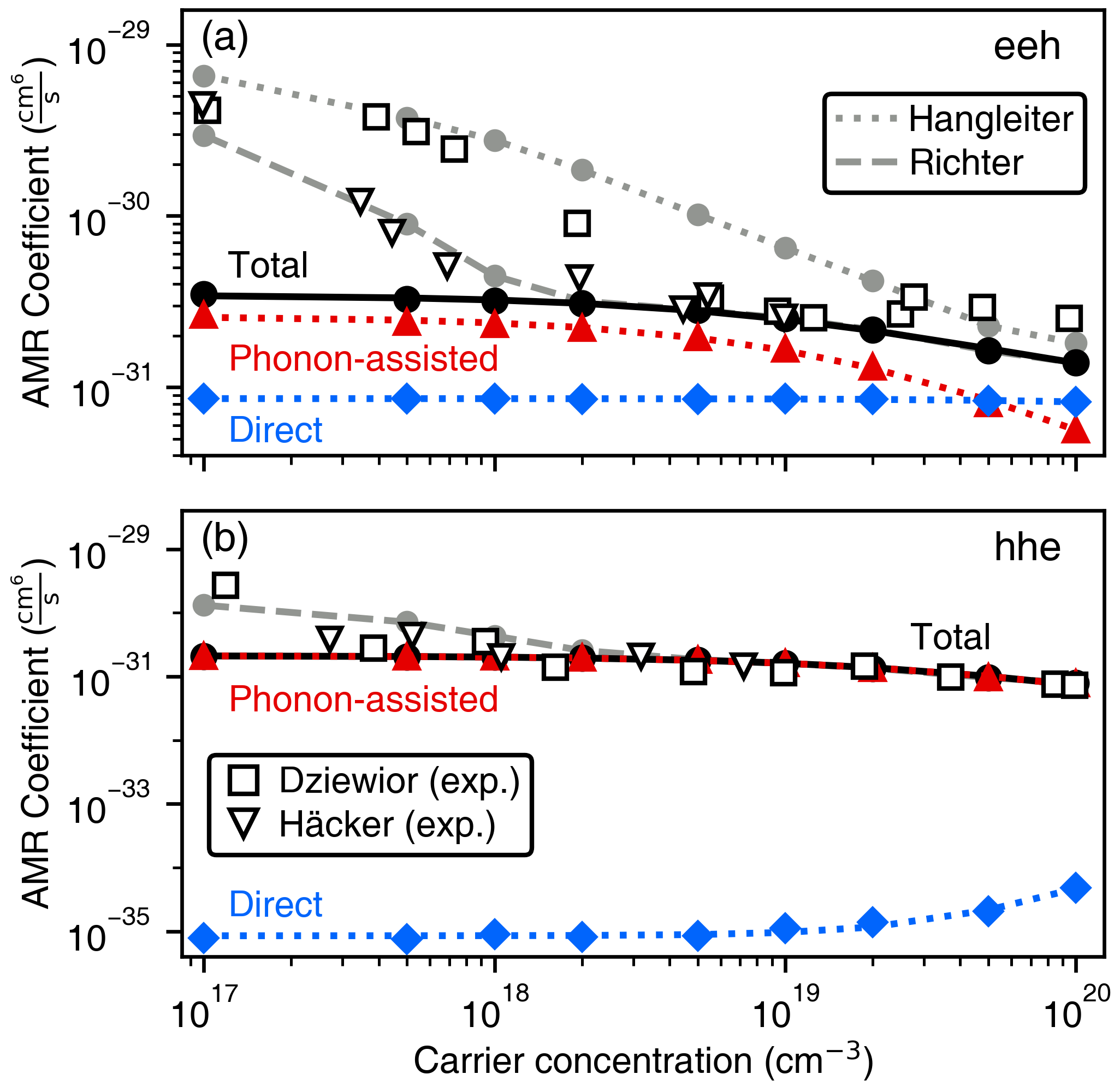}
\caption{AMR coefficient (\(C\)) as a function of carrier concentration for (a) the \(eeh\) process and (b) the \(hhe\) process at 300 K. We decompose the total rate (black circles) into contributions from the phonon-assisted (red triangles) and direct (blue diamonds) processes. The curves are generated by our model (Eq. S4). We also compare our results to experimental measurements,\cite{Dziewior1997,Hacker1994} and approximate Coulomb-enhancement effects using existing models.\cite{Hangleiter1990,Richter2012a}}
\label{fig:CvN}
\end{figure}

The dependence of the direct and \pa processes on carrier concentration is shown in Fig. \ref{fig:CvN}. Overall, the AMR coefficient remains approximately constant, though it does decrease at high carrier concentrations. This behavior is due to the fact that the Coulomb interaction is short range for direct AMR, but both short- and long-range interactions play a role for \pa AMR. Thus, the increased screening at higher carrier concentrations affects the \pa rate more strongly. On the other hand, at lower carrier concentrations weaker screening enables electron-hole interactions to increase the local carrier concentrations of holes (electrons) around electrons (holes), and therefore increase the rate of AMR recombination, an effect known as Coulomb enhancement.\cite{Richter2012a,Hangleiter1990,Altermatt1997} Including these many-body effects is outside the scope of our work, but Hangleiter and H\"{a}cker and Richter et al. have developed Coulomb-enhancement factors which we apply to our results in Fig. \ref{fig:CvN} to estimate such effects.\cite{Hangleiter1990,Richter2012a} We see that both correction factors recover the correct trend, validating that Coulomb-enhancement is responsible for the discrepancy between our calculations and measurements at lower carrier concentrations. At higher carrier concentrations, increased screening neutralizes the Coulomb-enhancement effects. We provide a physically informed model for \(C(n,p)\) in the Supplemental Material,\cite{supplemental} including Refs. \citenum{David2010} and \citenum{McAllister2018}.

\begin{figure}[ht]
\includegraphics{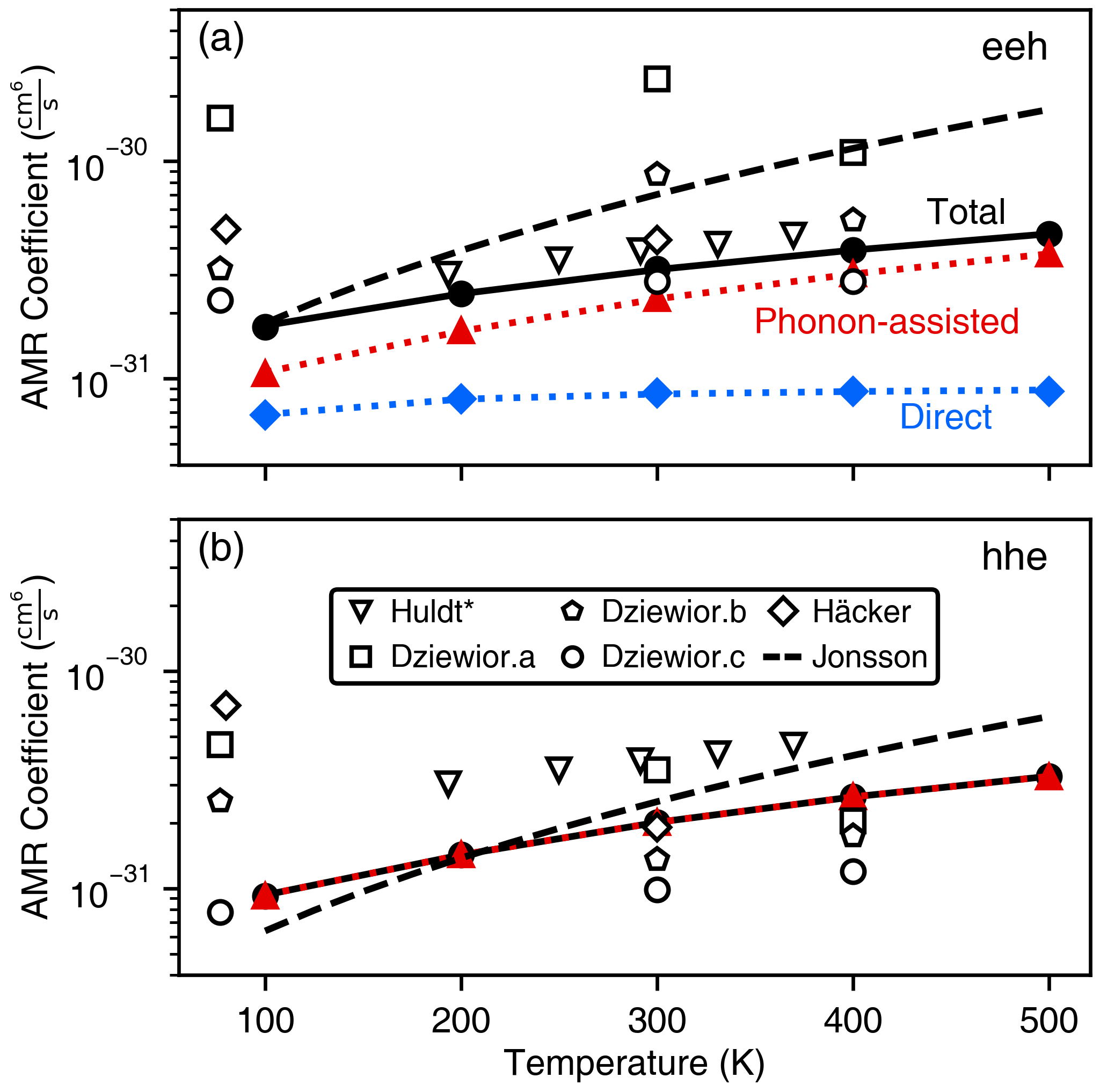}
\caption{AMR coefficient as a function of temperature for (a) the \(eeh\) process and (b) the \(hhe\) process at a carrier concentration of \oee. We decompose the total rate (black circles) into contributions from the phonon-assisted (red triangles) and direct (blue diamonds) processes. The curves are generated by our model (Eq. S5). Experimental data points are also included for comparison, though we note that not all points are at carrier concentrations of exactly \oee.\cite{Huldt1979,Dziewior1997,Hacker1994,Jonsson1997}}
\label{fig:CvT}
\end{figure}

We also examined the effect of temperature on the direct and \pa processes at a fixed carrier concentration of \oee (Fig. \ref{fig:CvT}). We find that temperature has a negligible effect on the direct AMR coefficient but the \pa process is more sensitive to temperature and follows the Bose-Einstein distribution, indicating that an increasing phonon population is the primary driver of this temperature dependence. While there is sparse experimental data investigating the temperature dependence of AMR at these carrier concentrations, we do find reasonable agreement between our calculations and the range of available experiments. We also formulate a physically motivated parameterization for \(C(T)\), which is discussed in the Supplemental Material and includes Ref. \citenum{McAllister2015}.\cite{supplemental}

\begin{figure*}[ht]
\includegraphics{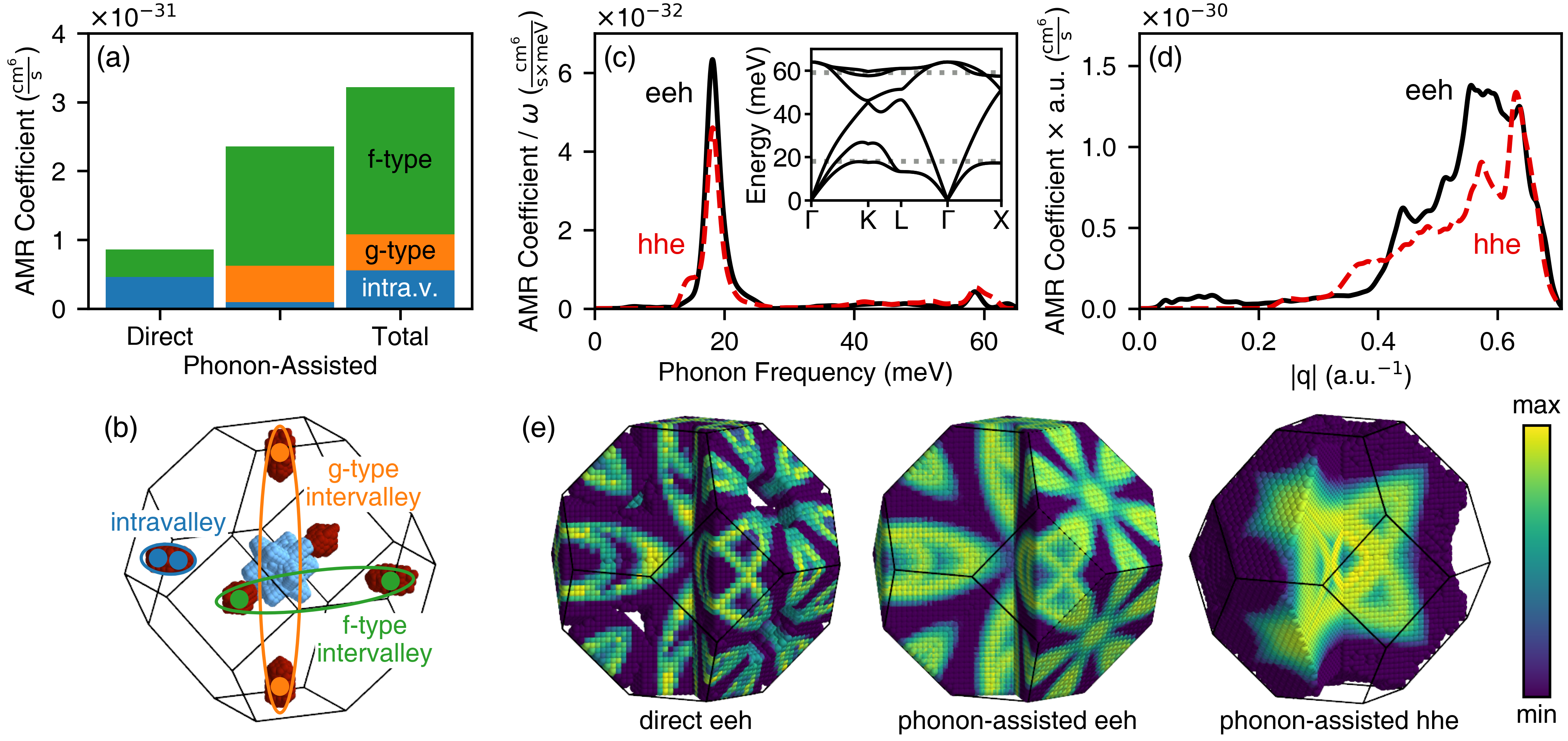}
\caption{Analysis of different contributions to the overall AMR rate. (a) Relative importance of the three different initial valley arrangements for electrons in the \(eeh\) process, which are illustrated in (b) with the \textit{f}-type arrangement contributing most strongly. The strength of \pa AMR for \(eeh\) (solid black) and \(hhe\) (red dash) processes as a function of phonon energy (c) and wave vector magnitude (d), where the strongest peaks are associated with TA phonons, highlighted in the inset phonon dispersion. (e) The distribution of excited carrier states throughout the first Brillioun zone for the direct and \pa \(eeh\) and \pa \(hhe\) processes, with slices removed to show the internal structure.}
\label{fig:analysis}
\end{figure*}

By leveraging our first-principles methodology we are able to decompose the overall AMR rate into distinct valley and phonon contributions. In the \(eeh\) process, there are three unique possibilities for the starting valley arrangements of the participating electrons: both electrons originating in the same valley (intravalley), electrons originating in opposite valleys (\textit{g}-type), or electrons originating in perpendicular valleys (\textit{f}-type), shown in Fig. \ref{fig:analysis}b (although we follow the same notation as phonon-scattering processes, we emphasize that these descriptions do not refer to scattering but rather to the initial valley arrangement of the two electrons during \(eeh\) AMR). The \textit{f}-type and intravalley terms are nearly equal for direct \(eeh\) AMR (with negligible \textit{g}-type contribution), consistent with the findings of Laks et al.\cite{Laks1990}, while the \textit{f}-type arrangement dominates for \pa recombination, and therefore for the overall rate (Fig. \ref{fig:analysis}a). This finding demonstrates a pathway for modulating the AMR rate via strain engineering: applying biaxial strain changes the relative energies of the different conduction band valleys and therefore the distribution of electrons, tuning the viability of different valley arrangements. Indeed, using strain to modulate the carrier occupation in bulk semiconductors is a technique that has been applied to engineer a variety of material properties, including the carrier mobility.\cite{Yu2008,Bushick2020,Ponce2019a}

We also investigate \pa AMR as a function of the participating phonon energies and momenta. Decomposing the total coefficient by phonon frequency (Fig. \ref{fig:analysis}c), we observe two distinct peaks around 18 (short wavelength TA modes) and 59 meV (primarily TO modes). The peak at 18 meV accounts for \({\sim}78\%\) of the \(eeh\) and \({\sim}66\%\) of the \(hhe\) AMR coefficient. These fractions indicate that while phonons across the energy spectrum play a role in \pa AMR, the zone-edge acoustic phonons dominate. Furthermore, the phonon emission process accounts for the majority of both \(eeh\) (74\%) and \(hhe\) (67\%) \pa AMR processes, indicating that it is not possible to freeze out the \pa AMR mechanism in silicon even at cryogenic temperatures. We also analyze the contributions of different phonon wave vectors to the overall \pa AMR rate (Fig. \ref{fig:analysis}d). We observe that at long wavelengths (\textbar q\textbar \(<\) 0.2 a.u.\textsuperscript{-1}) \(hhe\) \pa AMR is forbidden (by momentum conservation) while \(eeh\) \pa AMR is extremely weak (only 3\% of the total). This result justifies approximating the carrier wave function of wave vector \textbf{k} by those at the nearest band extrema, as any small deviations from the band extrema are negligible compared to the large wave vectors \(\mathbf{q}\) that dominate the \pa AMR process. Further analysis and visualization of the contributing phonon modes are found in the Supplemental Material.\cite{supplemental} 

Finally, we examine the distribution of the final states of excited electrons and holes. Plotting this distribution over the BZ provides a clear example of why the \pa process is important in \(eeh\) AMR. The distributions of the direct and \pa processes are nearly identical (Fig. \ref{fig:analysis}e), with the exception being states near the L point. These states contribute strongly to the overall AMR rate yet are inaccessible without the additional momentum provided by phonons. While the excited hole isosurface has similar topology to the hole Fermi surface (as in Fig. \ref{fig:analysis}b), there are also new features that emerge that are unique to the high-energy holes, such as the lobes towards the L points. In both \(eeh\) and \(hhe\) cases, this analysis also illustrates the large momentum transfer occurring in the Coulomb interaction, which reinforces the need for our first-principles atomistic treatment of AMR as methods such as \(k\cdot{p}\) do not describe the bands accurately far from the band extrema.

In summary, we investigated the direct and \pa AMR rate in silicon from first principles. Our calculations are in excellent agreement with experimental values and demonstrate the importance of the \pa AMR process in both \textit{n}-type (\(eeh\) dominant) and \textit{p}-type (\(hhe\) dominant) silicon, answering the long-standing question about the role of phonons in AMR in silicon. Our analysis shows that short wavelength phonons (primarily acoustic) dominate the \pa AMR mechanism. The large momentum transfer involved in both the Coulomb and electron-phonon scattering processes underscores the need for a first-principles atomistic treatment of AMR in silicon. We further propose a potential pathway for modulating the AMR rate by tuning the carrier occupations of different conduction band valleys via strain engineering. Our methodology elucidates the microscopic origins of AMR in silicon and paves the way for unprecedented scientific understanding and engineering of this fundamental recombination mechanism in other technologically important direct- and indirect-gap semiconductors.

\begin{acknowledgments}
We thank Chris Van de Walle and David Young for helpful discussions. The work is supported as part of the Computational Materials Sciences Program funded by the U.S. Department of Energy, Office of Science, Basic Energy Sciences under Award No. DE-SC0020129. This work used resources of the National Energy Research Scientific Computing (NERSC) Center, a DOE Office of Science User Facility supported under Contract No. DE-AC02–05CH11231. K.B. acknowledges the support of the U.S. Department of Energy, Office of Science, Office of Advanced Scientific Computing Research, Department of Energy Computational Science Graduate Fellowship under Award No. DE-SC0020347. 
\end{acknowledgments}
\bibliography{amr.bib}

\clearpage
\onecolumngrid
\beginsupplement
\appendix

\begin{center}
    \large{\textbf{Supplemental Material for Phonon-assisted \am Recombination in Silicon from First Principles}}
    
    \vspace{3mm}
    
    \normalsize{Kyle Bushick and Emmanouil Kioupakis}
    
    \small{\textit{Department of Materials Science and Engineering, University of Michigan}}
    
    \small{\textit{Ann Arbor, Michigan 48109}}
\end{center}
\bigskip

\beginsupplement

\begin{center}
\textbf{Summary of Perturbation Theory Treatment of AMR}
\end{center}

The AMR coefficient, \(C = \frac{R}{Vn^3}\), can be evaluated using perturbation theory and Fermi's golden rule.\cite{Kioupakis2015} The direct AMR rate is described by first-order perturbation theory with
\begin{align}
\begin{split}
     R_{d} &= 2\frac{2\pi}{\hbar}\sum_\mathbf{1234}f_\mathbf{1}f_\mathbf{2}(1-f_\mathbf{3})(1-f_\mathbf{4})\\&\times|M_{\mathbf{1234}}|^2\delta(\epsilon_\mathbf{1}+\epsilon_\mathbf{2}-\epsilon_\mathbf{3}-\epsilon_\mathbf{4}),
\end{split}
\label{eq:direct}
\end{align}
where the bold indices represent composite band and wave vector indices \([\mathbf{1}\equiv(n_1,\mathbf{k}_1)]\), \(f\) are the Fermi-Dirac occupation numbers, \(M_{\mathbf{1234}}\) is the screened Coulomb interaction between the carriers, and the \(\delta\) function enforces energy conservation. The screened Coulomb interaction includes both direct and exchange terms, defined as: 
\begin{equation}   
    M_{\mathbf{1234}}^d\equiv\braket{\psi_\mathbf{1}\psi_\mathbf{2}|W|\psi_\mathbf{3}\psi_\mathbf{4}},
\end{equation}
and 
\begin{equation}
    M_{\mathbf{1234}}^x\equiv\braket{\psi_\mathbf{1}\psi_\mathbf{2}|W|\psi_\mathbf{4}\psi_\mathbf{3}},
\end{equation}
with 
\begin{equation}
    \braket{\psi_\mathbf{1}\psi_\mathbf{2}|W|\psi_\mathbf{3}\psi_\mathbf{4}}=\iint d\bm{r}_1 d\bm{r}_2 \psi_\mathbf{1}^*(\bm{r}_1) \psi_\mathbf{2}^*(\bm{r}_2) W(\bm{r}_1,\bm{r}_2) \psi_\mathbf{3}(\bm{r}_1) \psi_\mathbf{4}(\bm{r}_2), 
\end{equation}
and 
\begin{equation}
    W(\bm{r}_1,\bm{r}_2) = \frac{1}{V}\sum_{\bm{q}}\frac{1}{\varepsilon(\bm{q})}\frac{4{\pi}e^2}{\bm{q}^2+\lambda^2}e^{i\bm{q}\cdot(\bm{r}_1-\bm{r}_2)}, 
\end{equation}
where \(V\) is the volume of the unit cell, \(\bm{q}\) is the transferred momentum, \(\varepsilon(\bm{q})\) is a model dielectric function from Cappellini which has been shown to work for many semiconductors,\cite{Cappellini1993} and \(\lambda\) is the screening length given by our screening model. In practice we calculate the matrix element by computing two overlap integrals of the wave functions:
\begin{equation}
    I_{\alpha,\beta}(\bm{G})=\int_{cell} u_{\alpha}^*(\bm{r})u_{\beta}(\bm{r})e^{i\bm{G}\cdot\bm{r}}d\bm{r}
\end{equation}
where \(u_{\alpha}(\bm{r})\) and \(u_{\beta}(\bm{r})\) are the periodic parts of the wave functions and \(\bm{G}\) are lattice vectors. The interested reader can find additional derivation details and discussion in Ref. \citenum{Kioupakis2015}.

The \pa AMR rate is given by second-order perturbation theory by
\begin{align}
\begin{split}
     R_{pa} &= 2\frac{2\pi}{\hbar}\sum_{\mathbf{1234};\nu\mathbf{q}}f_\mathbf{1}f_\mathbf{2}(1-f_\mathbf{3})(1-f_\mathbf{4})(n_{\nu\mathbf{q}}+\frac{1}{2}\pm\frac{1}{2})\\&\times|\tilde{M}_{\mathbf{1234};\nu\mathbf{q}}|^2\delta(\epsilon_\mathbf{1}+\epsilon_\mathbf{2}-\epsilon_\mathbf{3}-\epsilon_\mathbf{4}\mp\hbar\omega_{\nu\mathbf{q}}),
\end{split}
\label{eq:indirect}
\end{align}
where the summation includes phonons of mode \(\nu\) and wave vector \(\mathbf{q}\), \(n_{\nu\mathbf{q}}\) is the Bose-Einstein occupation number of the phonon, and the upper (lower) sign corresponds to the phonon emission (absorption) process. \(\tilde{M}_{\mathbf{1234},\nu\mathbf{q}}\) is the generalized matrix element for the \pa \am process and includes all combinations of the electron-phonon and screened Coulomb interaction Hamiltonians. While Ref. \citenum{Kioupakis2015} describes these equations in greater detail, the general form of the matrix element contributing to \(\tilde{M}\) is 
\begin{equation}
     \tilde{M}^1_{\mathbf{1234},\nu\mathbf{q}} = \sum_\mathbf{m}\frac{g_{\mathbf{1m};\nu}M^d_\mathbf{m234}}{\epsilon_\mathbf{m}-\epsilon_\mathbf{1}\pm\hbar\omega_{\nu\mathbf{q}}+i\eta},
\label{eq:me}
\end{equation}
where \(g_{\mathbf{1m};\nu}\) is the electron-phonon coupling matrix element for an electron between initial state \(\mathbf{1}\) and final state \(\mathbf{m}\equiv(m_1,\mathbf{k}_1+\mathbf{q})\) mediated by phonon mode \(\nu\) with wave vector \(\mathbf{q}\) and is obtained from Quantum ESPRESSO, \(M^d_\mathbf{m234}\) is the direct term of the screened Coulomb matrix element, and \(i\eta\) is a small imaginary term to prevent singularities in the denominator. As seen in Eq. 10 of Ref. \citenum{Kioupakis2015}, there are four direct and four exchange matrix elements involved in the \pa AMR process. 

In transforming Eq. \ref{eq:indirect} into Eq. 1, we note that the meaning of the composite band indices change. \(\mathbf{123}\) now represent discrete sums over the degenerate bands and valleys at the valence and conduction band extrema. These combine with \(\mathbf{q}\) sampled over the irreducible BZ to define the \textbf{k}-point in \(\mathbf{4}\).

\begin{center}
\textbf{LDA and GW Bandstructure Calculations}
\end{center}

Our LDA calculations are preformed on a relaxed structure with a lattice parameter of 5.379 \AA, which is within 1\% of the experimental lattice parameter of 5.431 \AA.\cite{Okada1984} We use a 40 Ry plane wave cutoff for our norm-conserving pseudopotential and an 8\(\times\)8\(\times\)8 Monkhorst-Pack Brillouin-zone (BZ) sampling grid. While these parameters are over-converged, they are necessary to obtain an accurate interpolation using the maximally localized Wannier function method.

For the G\textsubscript{0}W\textsubscript{0} (GW) calculations, we use an 8\(\times\)8\(\times\)8 BZ sampling mesh, a 35 Ry dielectric-matrix cutoff (which sets the maximum kinetic energy of G-vectors included in the dielectric matrix), a 30 Ry screened Coulomb cutoff (which sets the maximum kinetic energy of G-vectors included in the screened Coulomb interaction to obtain the quasiparticle self-energies), and sums over 700 and 660 bands for the dielectric-function and self-energy sums over unoccupied states, respectively. We use the Generalized Plasmon Pole model of Hybertsen and Louie \cite{Hybertsen1986a} and the static remainder correction.\cite{Deslippe2013}

\begin{figure}[ht]
\includegraphics{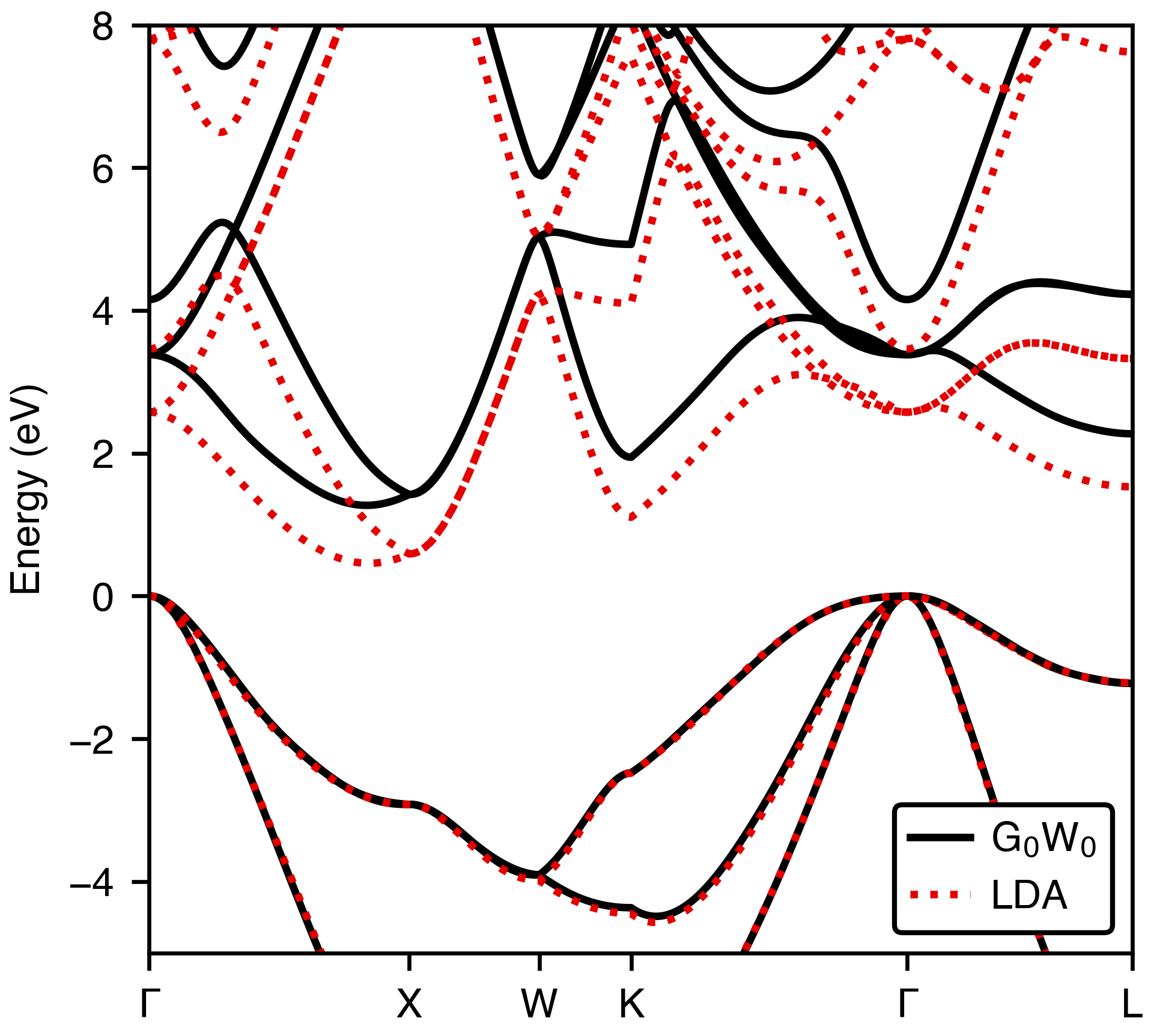}
\caption{G\textsubscript{0}W\textsubscript{0} (solid black) and LDA (dashed red) band structures of silicon, with energy referenced to the G\textsubscript{0}W\textsubscript{0} valence band maximum.}
\label{fig:sup_bands}
\end{figure}

In Fig. \ref{fig:sup_bands}, we show both the GW and LDA band structures, referencing both to their respective valence band maximum values. The LDA band gap is 0.458 eV, while the GW quasiparticle corrections open the gap to 1.27 eV. However, while the GW band gap is more accurate than LDA, the actual band gap of silicon at 300 K is 1.12 eV.\cite{Bludau1974} Given this, we implement a rigid shift to the GW eigenvalues in order to calculate the AMR rates at adjustable band gap values. The motivation for such a shift is threefold: (1) it allows us to match the experimental band gap while (2) also assessing our calculation convergence with respect to the BZ sampling grid as well as (3) assessing the sensitivity of the AMR coefficient to the band gap value. Unless mentioned explicitly, all AMR coefficient data reported throughout this work are for a band gap of 1.12 eV. 

\begin{figure}[!ht]
\includegraphics[width=6.5in]{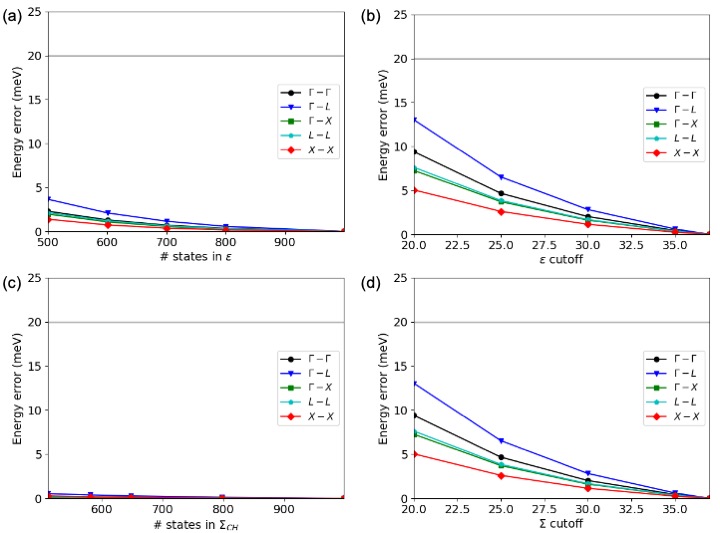}
\caption{Convergence parameters for the GW calculation. (a) shows the convergence of energy error as a function of the sum over bands for constructing the dielectric-function (b) shows the convergence as a function of the dielectric-matrix cutoff (c) shows the convergence as a function of the sum over bands for the self-energy, and (d) shows the convergence as a function of the screened Coulomb cutoff. The different curves show energy differences between different high symmetry points across the band gap, referenced to the most converged value at \(\Gamma-\Gamma\).}
\label{fig:gw_conv}
\end{figure}

While we report the final parameters used in the GW calculation above, Fig. \ref{fig:gw_conv} shows the convergence tests carried out. The plots show the energy error between a handful of high-symmetry points. Each parameter is tested independently, with the other three set to the greatest value while the fourth is varied. For the number of bands for constructing the dielectric-function we test 500, 601, 700, 800, and 999 as our values (Fig. \ref{fig:gw_conv}a). For the number of bands used in the self-energy sum we test 512, 582, 648, 798, and 998 as our values (Fig. \ref{fig:gw_conv}c). These values are slightly different from one another because of the degeneracy restrictions. For both the dielectric-matrix and screened Coulomb cutoff energies we test 20, 25, 30, 35, and 37 Ry (Fig. \ref{fig:gw_conv}b and Fig. \ref{fig:gw_conv}d). Our GW calculations used for our subsequent calculations are thus converged to within 15 meV.

\begin{center}
\textbf{Convergence of Fermi Energies}
\end{center}

\begin{figure}[h]
\includegraphics{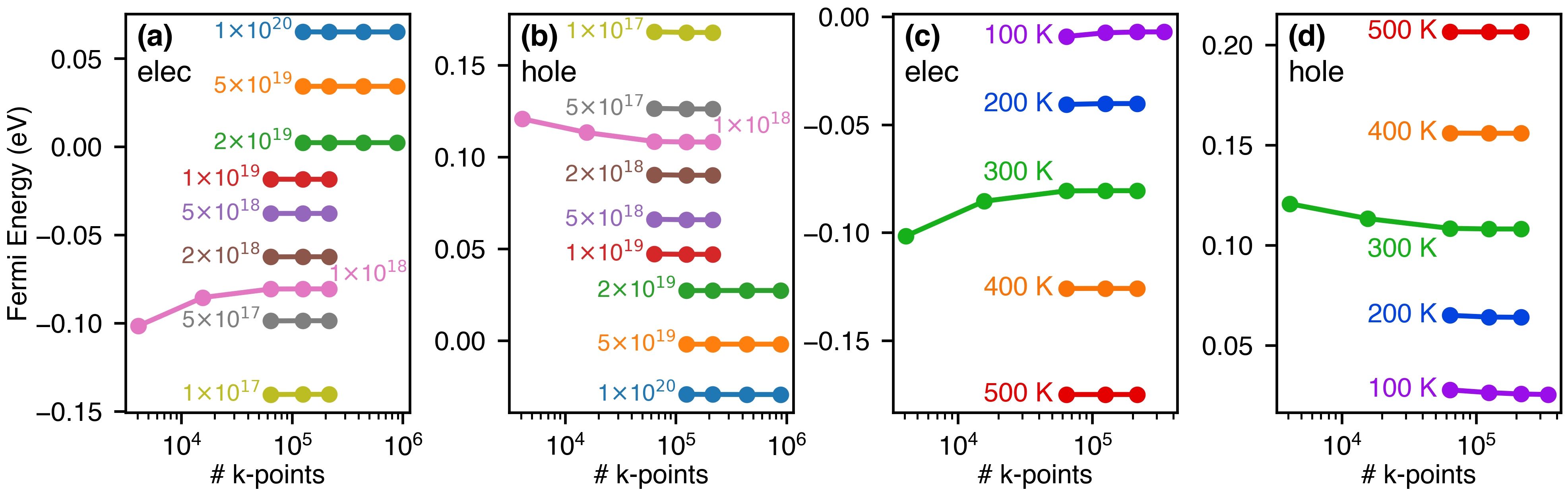}
\caption{Convergence of Fermi energies for electrons and holes (referenced to the corresponding band extrema) with respect to carrier concentration (a,b) and temperature (c,d) as a function of the BZ sampling. When the carrier concentration is varied, the temperature is fixed at 300 K, while when the temperature is varied the carrier concentration is fixed at \oee.}
\label{fig:fermi_conv}
\end{figure}

\begin{figure}[h]
\includegraphics{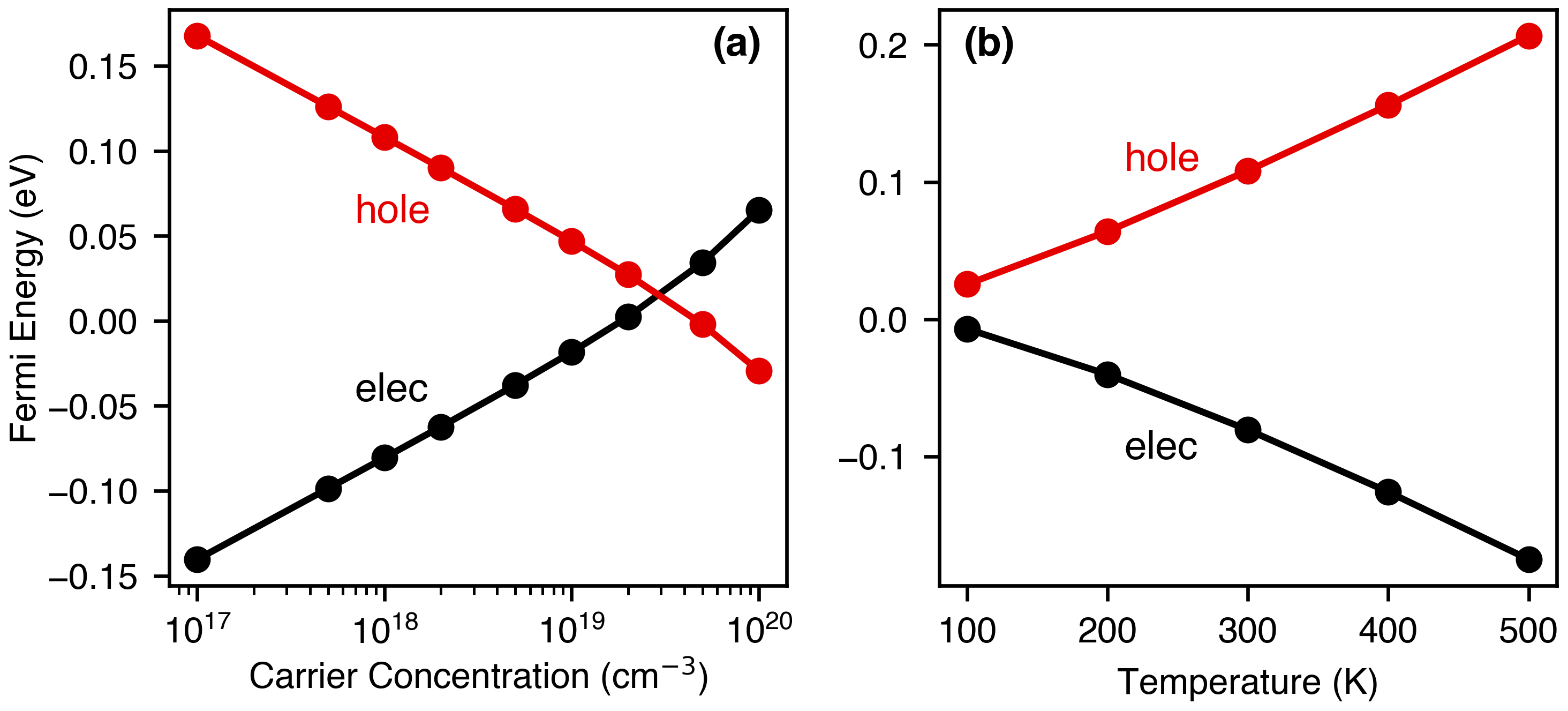}
\caption{Fermi energy trends for both electrons and holes (referenced to the corresponding band extrema) as a function of carrier concentration (a) and temperature (b).}
\label{fig:fermi_final}
\end{figure}

One of the primary inputs to our AMR calculations is the free-carrier Fermi energy, which is determined iteratively using the bisection method for a given carrier concentration and temperature. Note that we define the Fermi energy relative to the respective band edge; the Fermi energy for electrons (holes) is referenced to the conduction (valence) band minimum (maximum). Thus, non-degenerate Fermi energies are negative (positive) for electrons (holes). As part of this calculation process, the Fermi energy must be converged as a function of the BZ sampling grid. In Fig. \ref{fig:fermi_conv} we show the convergence tests, which were first done more extensively for the case of the 300 K and \oee condition, with more targeted tests used for the other conditions. From these converged values, we show the trends of the converged values versus carrier concentration and temperature in Fig. \ref{fig:fermi_final}. From this plot, it is clear that while we are investigating a range of conditions, we are primarily in the non-degenerate doping regime, only crossing into the degenerate doping regime at 300 K for carrier concentrations above \(2\times10^{19}\) \cmc, while varying temperature at \oee remains non-degenerate, though 100 K approaches the transition. 

\clearpage

\begin{center}
\textbf{Convergence of \am Recombination Coefficients}
\end{center} 

\begin{figure}[h]
\includegraphics{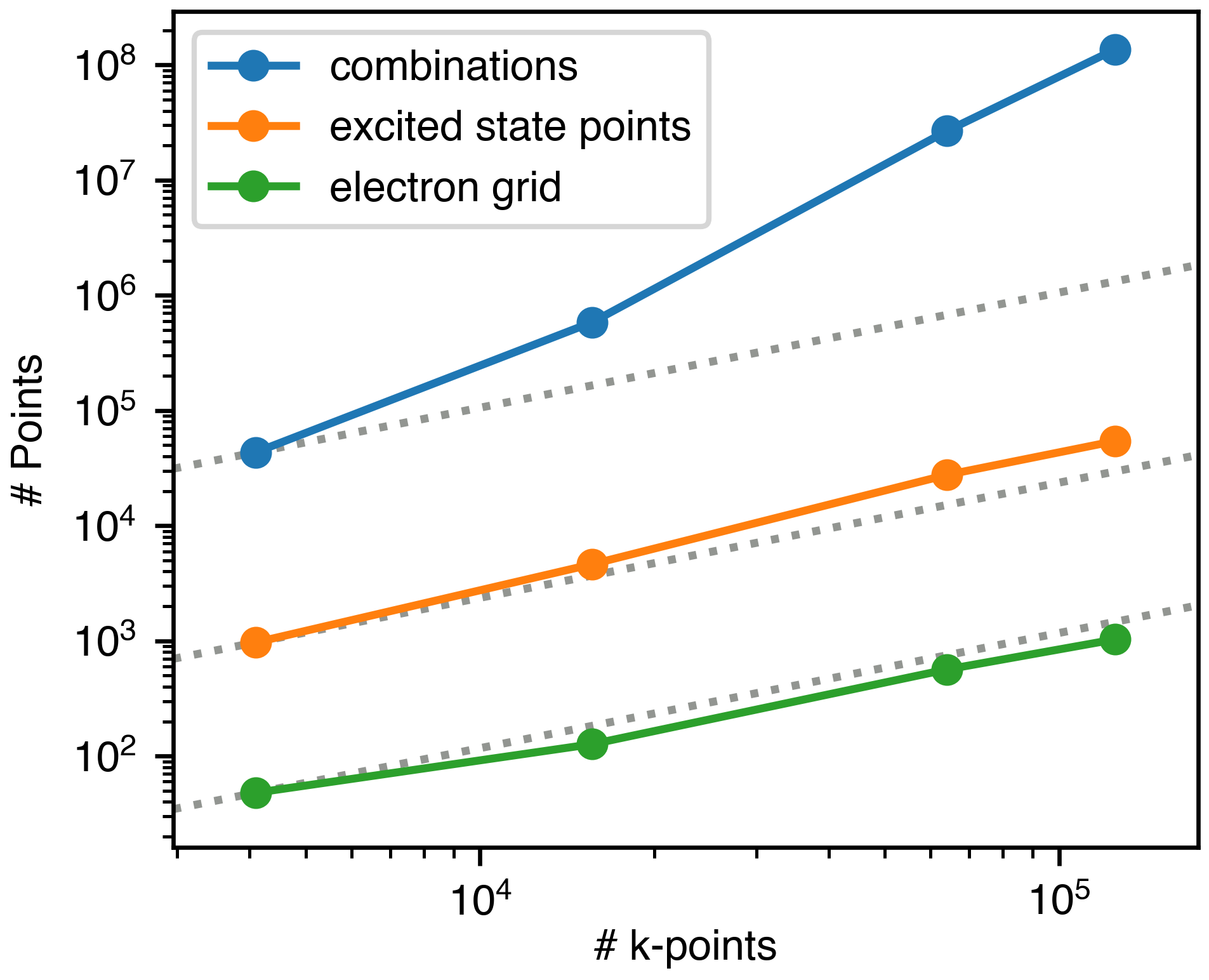}
\caption{Scaling of the number of points as a function of BZ sampling grid for the direct \(eeh\) \am process. The grey lines are linear sloped curves, and are provided as guides to the eye.}
\label{fig:scaling}
\end{figure}

Following the determination of Fermi energies, the BZ sampling grid must be converged for the AMR coefficient calculation itself. For the direct \am calculations, the convergence process entails using increasingly fine sampling grids for the electrons and holes. The large computational cost of this method comes from the scaling of the problem as a function of grid size. As shown in Fig. \ref{fig:scaling} using the \(eeh\) process as an example, the number of points sampled for the electron grid, as well as the excited state points which are determined based on the electron and hole momenta, scale roughly linearly with the cube of the grid size, which is expected given that we are investigating a bulk property. However, the total number of combination (blue points) scales superlinearly with the cube of the grid size.

\begin{figure}[h]
\includegraphics{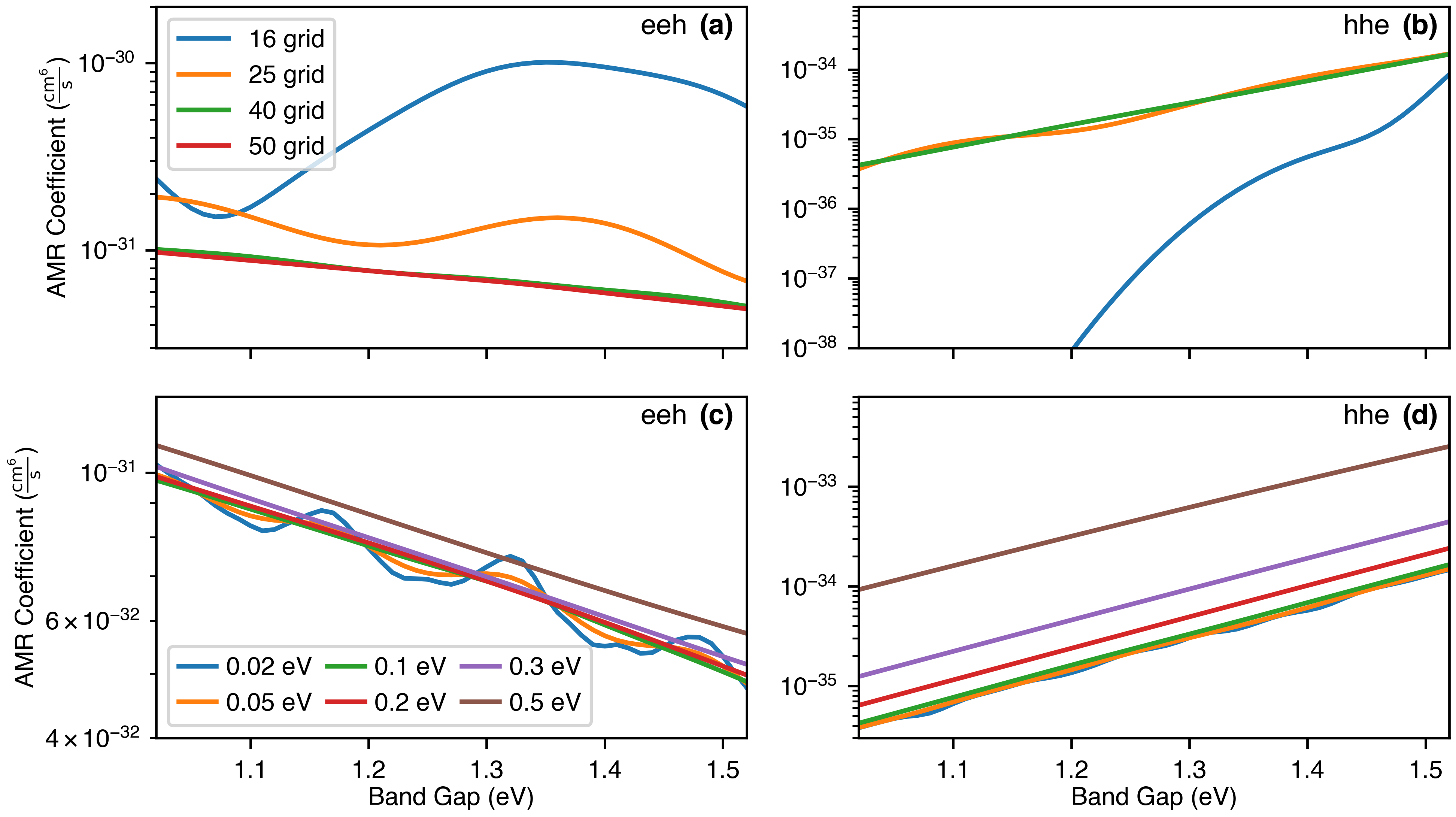}
\caption{Convergence testing for direct \am calculations at 300 K and \oee. (a) and (b) show the convergence as a function of sampling grid (for \(\sigma=0.1\) eV) for the \(eeh\) and \(hhe\) processes, respectively. (c) and (d) show the convergence as a function of \(\sigma\) for \(50\times50\times50\) (\(eeh\)) and \(40\times40\times40\) (\(hhe\)) BZ sampling grids, respectively.}
\label{fig:direct_conv}
\end{figure}

For a given direct \am calculation at a fixed grid size, two other parameters are varied: the rigid shift of the band gap and the delta function broadening parameter, \(\sigma\). By varying the band gap, we obtain a qualitative measure of the sensitivity of the AMR rate, while we use a broadening parameter \(\sigma\) to slightly relax the conservation requirements since we are constrained to a discrete sampling grid. If \(\sigma\) is too small, then the BZ sampling requires exceedingly fine grids, while too large of a \(\sigma\) value leads to an artificial smoothing and increase in the AMR rate, as too wide a range of states contribute to a given transition. The convergence testing for both the sampling grid and \(\sigma\) (for the \(T=300\) K, \(n=p=\) \oee conditions) are shown in Fig. \ref{fig:direct_conv}. Based on this testing, we use a \(\sigma\) value of 0.1 eV, while a \(50\times50\times50\) grid is used for the direct \(eeh\) process and a \(40\times40\times40\) grid is used for the direct \(hhe\) process. These parameters are chosen as they are large enough to remove the artificial oscillations due to under-sampling of the BZ while still being small enough to prevent artificial inflation of the AMR coefficient (for \(\sigma\)) and unnecessary computational cost (for the grid size). 

\begin{figure}[h]
\includegraphics{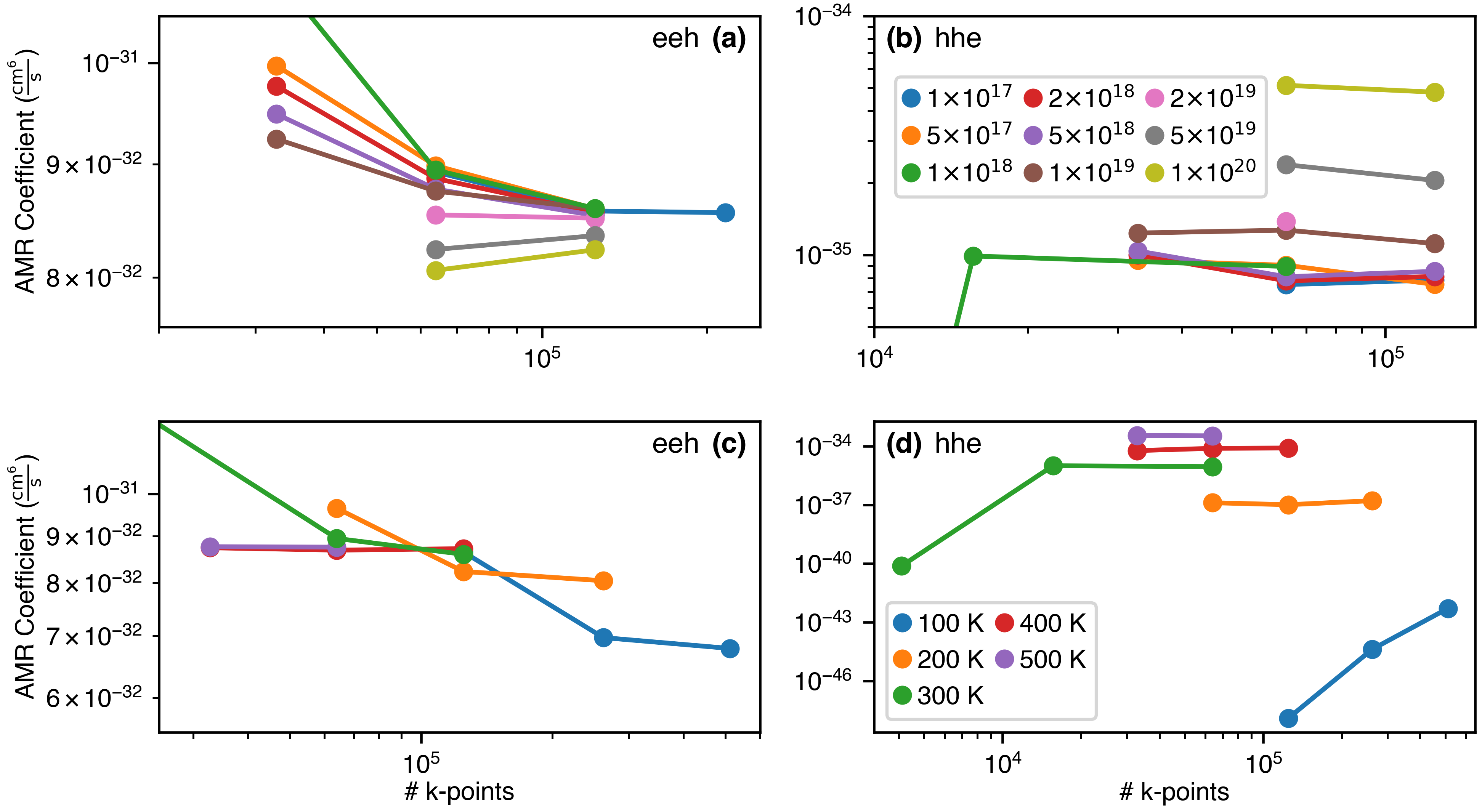}
\caption{Convergence testing for direct \am calculations at a variety of carrier concentrations and temperatures at the experimental band gap. (a) and (b) show the convergence of different carrier concentrations as a function of sampling grid for the \(eeh\) and \(hhe\) processes, respectively. (c) and (d) show the convergence of different temperatures in the same manner.}
\label{fig:direct_conc_temp_conv}
\end{figure}

Once we developed an understanding of the convergence behavior at one set of (\(T\), \(n\)) conditions, we conducted more abbreviated convergence testing at different carrier concentrations and temperatures. Fig. \ref{fig:direct_conc_temp_conv} shows the convergence as a function of BZ sampling grid at the experimental band gap and with \(\sigma=0.1\) eV. 

\begin{figure}[h]
\includegraphics[width=5.6in]{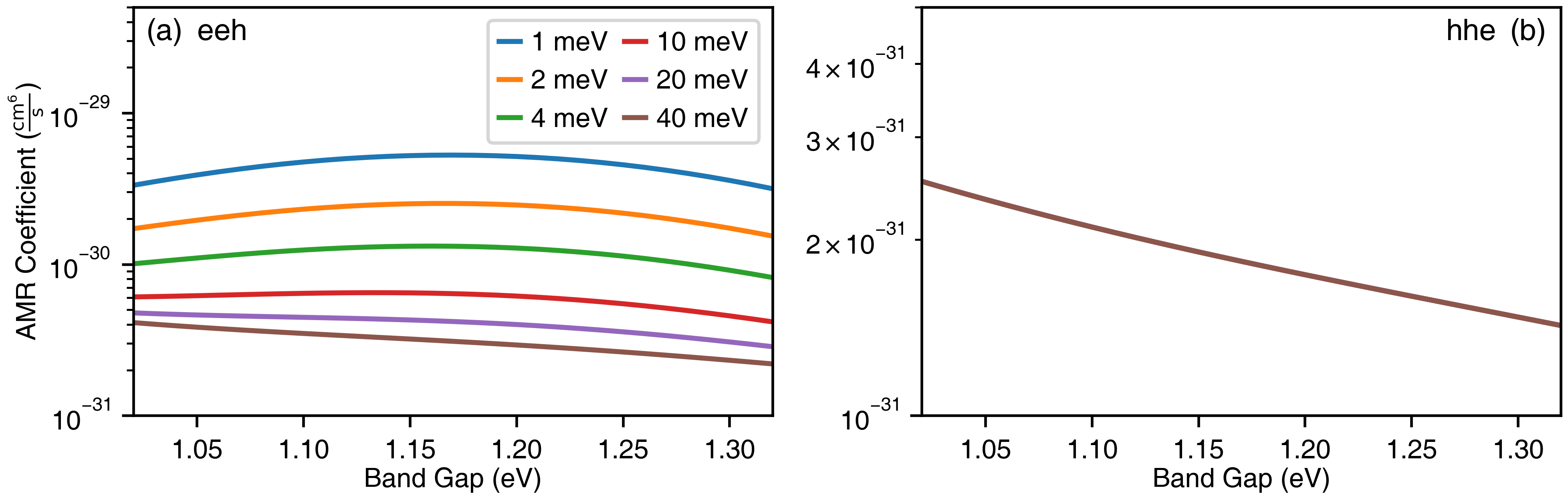}
\caption{Effect of varying the \(\eta\) value of the \(i\eta\) term in \pa \am calculations at 300 K and \oee. (a) and (b) show the \(eeh\) and \(hhe\) processes, respectively. All curves are from \(50\times50\times50\) BZ sampling grids and with \(\sigma=0.2\) eV.}
\label{fig:indirect_imeta}
\end{figure}

In the case of \pa \am calculations, the convergence process is different. For the direct \am calculations, we use finite grids to sample the states that participate in the \am recombination, but for the \pa process, we approximate the initial electron and hole wave functions with those at the valence band and conduction band extrema. With the initial points fixed, we instead use a sampling grid for the phonon wave vectors, which are otherwise completely unconstrained by momentum conservation and span the entire BZ. The combination of the band-extrema wave vectors and the given phonon wave vector defines the momentum and energy of the possible high-energy carrier state. Given the cost of phonon calculations using density functional perturbation theory, we use only the irreducible wedge of the BZ to sample the phonon wave vectors with appropriate weights, which enables much finer grid sizes than would otherwise be computationally tractable. Another parameter that must be considered for \pa \am is the \(i\eta\) term in the denominator of the \pa matrix element equation (Eq. \ref{eq:me}), which is included to counteract the numerical divergence that occurs if the energy difference in the denominator (i.e. \(\epsilon_\mathbf{m}-\epsilon_\mathbf{1}\pm\hbar\omega_{{\nu}q}\)) approaches zero. This divergence is a general feature of second-order perturbation theory that arises if first-order processes are possible, and can lead to an undefined matrix element if left uncorrected. As with \(\sigma\) used above for the delta function broadening, we want to obtain results which do not contain artifacts from this numerical treatment. For silicon, only the \(eeh\) process encounters such a divergence, since direct transitions are feasible considering momentum and energy conservation, which is seen by the dependence on the value of \(\eta\) in Fig. \ref{fig:indirect_imeta}a. Conversely, the curves are independent of \(\eta\) in Fig. \ref{fig:indirect_imeta}b, as no direct transitions are possible for the \(hhe\) process. We discuss our handling of the \(i\eta\) term in greater detail in the following section. 

\begin{figure}[h]
\includegraphics[width=5.6in]{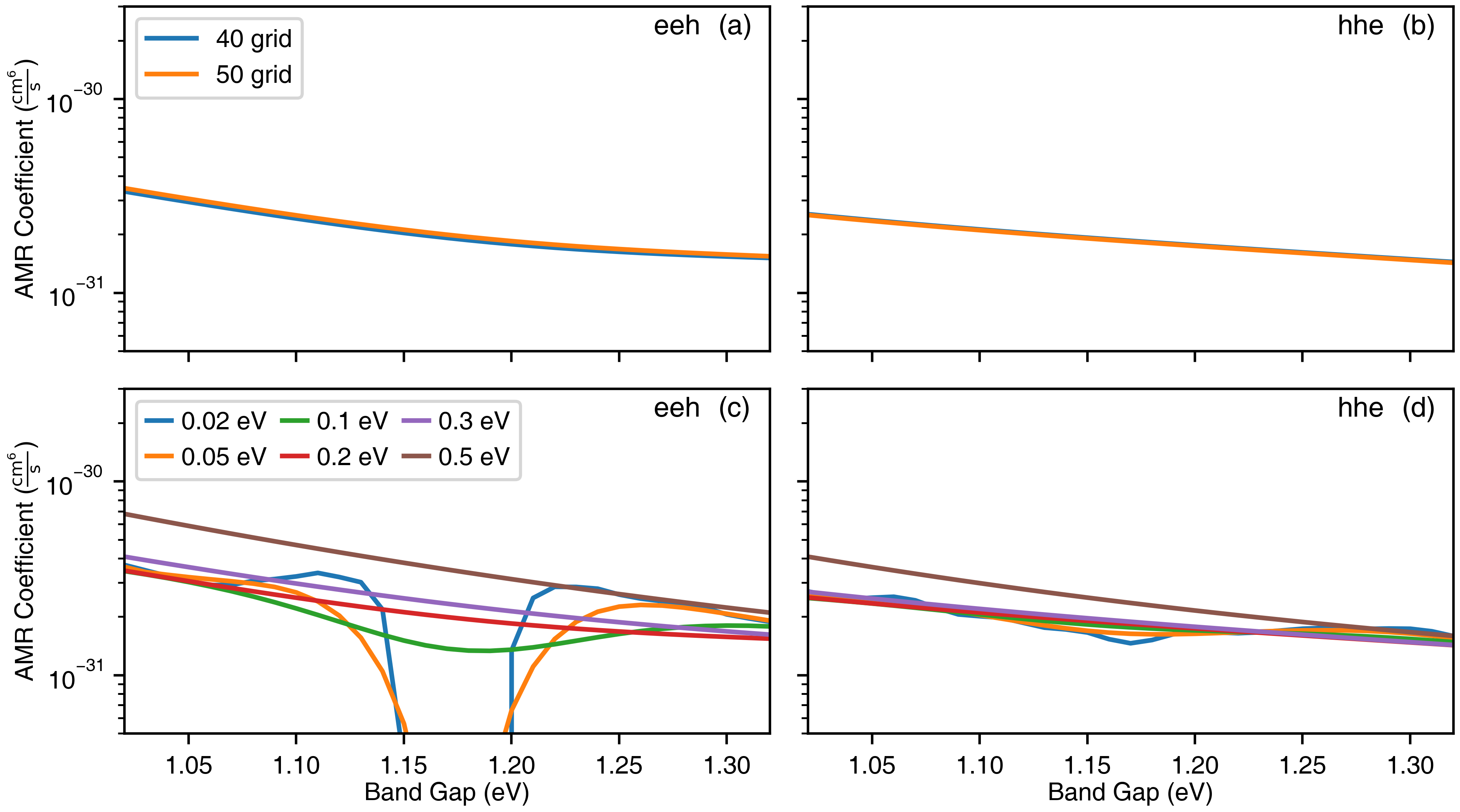}
\caption{Convergence testing for \pa \am calculations at 300 K and \oee. (a) and (b) show the convergence as a function of sampling grid (for \(\sigma=0.2\) eV) for the \(eeh\) and \(hhe\) processes, respectively. (c) and (d) show the convergence as a function of \(\sigma\) (for \(50\times50\times50\) BZ sampling grids) for the \(eeh\) and \(hhe\) processes, respectively. In all plots \(eeh\) data has been treated for the \(\eta\) dependence as discussed.}
\label{fig:indirect_conv}
\end{figure}

As in Fig. \ref{fig:direct_conv}, we show convergence corresponding to the sampling grid (now for phonon wave vectors) and delta function broadening, \(\sigma\), in Fig. \ref{fig:indirect_conv}. The values shown for \(eeh\) have been treated to deal with the \(\eta\) dependence in the manner presented in the next section. For the \pa calculations, \(\sigma=0.2\) eV proved to be the best balance, while sampling phonons from the irreducible wedge of the \(50\times50\times50\) grid demonstrated sufficient convergence. 

\clearpage

\begin{center}
\textbf{Phonon-assisted \textit{eeh} \am Divergence Correction}
\end{center}

\begin{figure}[h]
\includegraphics{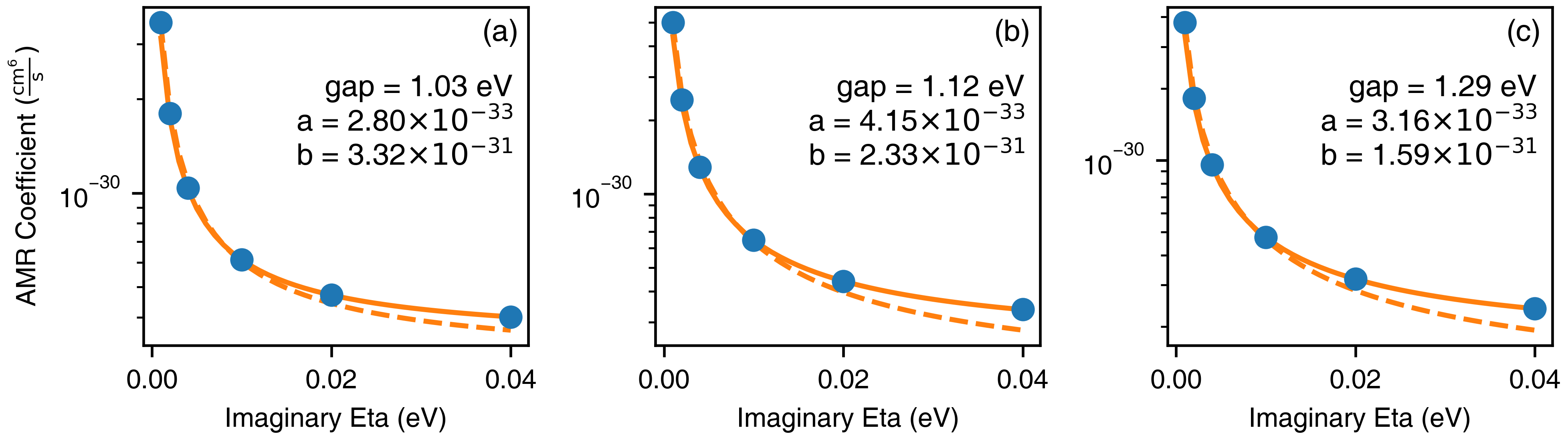}
\caption{Fit of \am coefficient versus \(\eta\) at three band gap values to \(f(x) = \frac{a}{x}+b\). The solid curve uses the three largest \(\eta\) values while the dashed curve uses all six points. The fit parameters shown on each plot are from the three-point fits.}
\label{fig:imeta_fit}
\end{figure}

As introduced above, \pa \am calculations include an extra term (\(i\eta\)) which prevents divergence near direct transitions. However, our results should be independent of the value of \(\eta\) {\textendash} as seen in Fig. \ref{fig:indirect_imeta}a this is not naturally the case. Since the divergence only occurs for the \(eeh\) process, the \(\eta\) value has no appreciable effect on the \(hhe\) results, and no further treatment is necessary. On the other hand, we use a fitting approach to find the \(\eta\)-independent \pa \(eeh\) AMR coefficient. As discussed by Brown et al., the expression for the matrix element in second-order perturbation theory can be written as a sum of an \(\eta\)-independent component and a component that follows an \(\eta^{-1}\) dependence, that is \(C \propto \frac{1}{\eta}+B\).\cite{Brown2016} While we calculate C at six different \(\eta\) values, the three smallest values (1, 2, and 4 meV) are likely under-converged at the grid sizes we employ in this work, ultimately leading to poor fits. We therefore fit \(C\) vs \(\eta\) to the three largest \(\eta\) values (10, 20, and 40 meV) to \(f(x) = \frac{a}{x}+b\), where \(b\) is the \(\eta\)-independent \pa AMR coefficient. This fit still captures the small \(\eta\) dependence well, giving us confidence in the fit which is no longer strongly influenced by the underconverged points with larger magnitude. In Fig. \ref{fig:imeta_fit}, we show both the full fit and three-point fit for \(C\) values below, at, and above the experimental band gap, demonstrating the robustness of the method.

\begin{center}
\textbf{Sensitivity to Effective Mass}
\end{center}

\begin{figure}[h]
\includegraphics{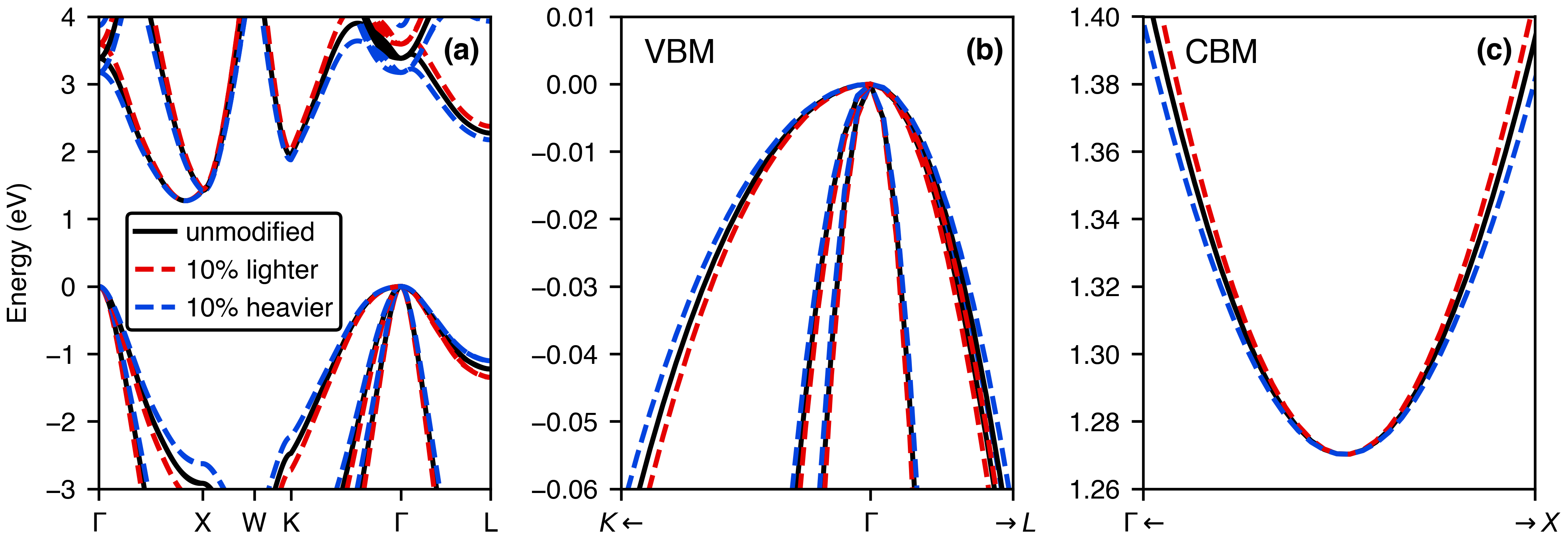}
\caption{(a) Band structure comparisons for the unmodified bands calculated with the G\textsubscript{0}W\textsubscript{0} method(black), 10\% lighter bands (red) and 10\% heavier bands (blue). (b) Zoomed view of the valence band maximum. (c) Zoomed view of the conduction band minimum.}
\label{fig:emh_bands}
\end{figure}

One of the arguments for using empirical or semi-empirical methods or pseudopotentials is that they allow one to fit to the experimental band gap and effective mass, two quantities that remain challenging to recover using first-principles approaches such as GW or hybrid functionals. We therefore seek to understand the impact of effective mass on the AMR rate for both direct and \pa mechanisms. To do this, we artificially modified the band curvature for two cases, such that we obtained effective masses that were 10\% lighter and 10\% heavier than the those obtained directly from our calculations. This is done by multiplying the eigenvalues by 1.1 (lighter) or 0.9 (heavier) and adjusting the absolute position so that the band extrema remain unchanged. The resultant band structures are shown in Fig. \ref{fig:emh_bands}. 

\begin{figure}[h]
\includegraphics{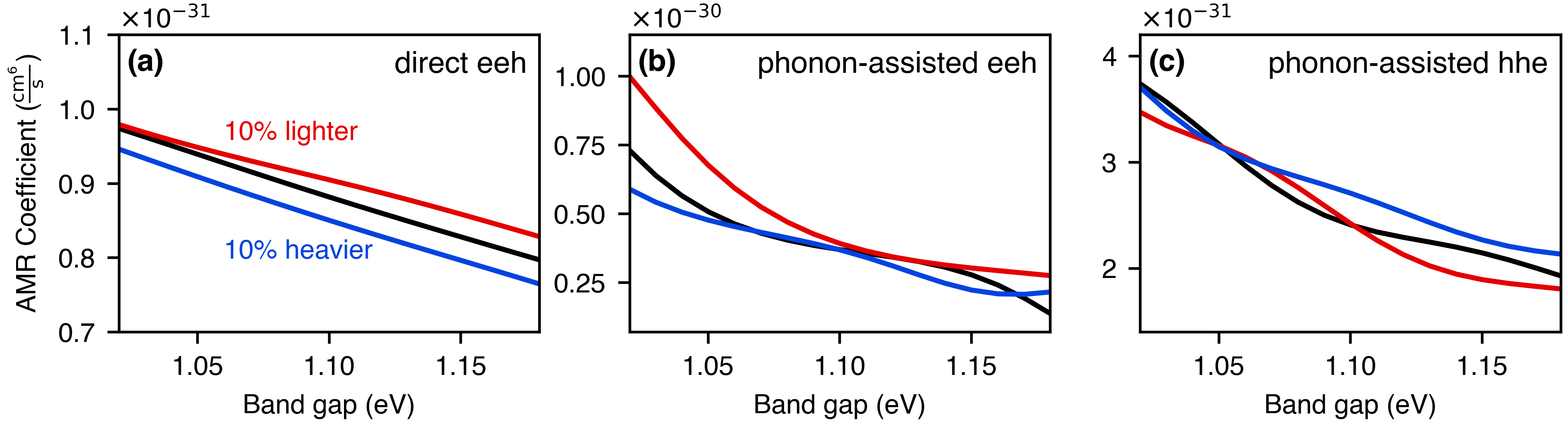}
\caption{(a) AMR coefficient versus band gap for the direct \(eeh\) process including the unmodified values (black) alongside those calculated from bands with 10\% lighter effective masses (red) as well as 10\% heavier effective masses (blue). The same is shown in (b) for the \pa \(eeh\) process and (c) for the \pa \(hhe\) process.}
\label{fig:emh_amr}
\end{figure}

We then calculate the direct and \pa AMR coefficients over a range of band gaps for both the \(eeh\) and \(hhe\) process (excluding direct \(hhe\)) on \(50\times50\times50\) grids. As we show in Fig. \ref{fig:emh_amr}, the effect of the varying effective mass is quite small with the AMR coefficients staying within the correct order of magnitude and largely differing by less than 15\% from the unmodified value. At the experimental band gap, for instance, the maximum deviation is within 11\%. Thus, while using a different effective mass may result in a slightly different reported value, it does not change our qualitative conclusions.

\clearpage

\begin{center}
\textbf{Physically Informed Models of the \am Recombination Coefficient}
\end{center}

While past empirical parameterizations of the AMR coefficient of silicon have not been able to isolate different contributions to the overall AMR coefficient, they have still offered great utility to the scientific community and for device engineering. To this end, we construct two physically informed models for the AMR coefficient as a function of both carrier concentration and temperature using our calculation results. Our model for the dependence of the AMR coefficient on the carrier concentration at 300 K is based on the effects of phase-space filling, increased carrier screening, and access to more transitions as the carrier concentration increases:\cite{David2010,McAllister2018} 
\begin{equation}
    C(n,p) = \frac{C_{dir}^{eeh}}{1+(\frac{n}{n^*_{dir}})^\alpha} + C_{dir}^{hhe} \biggl(1- \frac{C_{dir}^{hhe*}}{1+(\frac{p}{p^*_{dir}})^\beta}\biggr) + \frac{C_{pa}^{eeh}}{1+(\frac{n}{n^*_{pa}})^\gamma} + \frac{C_{pa}^{hhe}}{1+(\frac{p}{p^*_{pa}})^\delta},
    \label{eq:CvN}
\end{equation}
with the equation parameters listed in Table \ref{tab:CvN}. Though this model does not include the Coulomb-enhancement factor, one can readily apply any of the models presented in the literature.\cite{Hangleiter1990,Richter2012a,Black2022}


\begin{table}[h]
\centering
\begin{tabular}{ccc}
Parameter          & Value                   & Units \\ \hline
\(C_{dir}^{eeh}\)  & \(8.59\times 10^{-32}\) & cm\(^6\)s\(^{-1}\) \\
\(n^*_{dir}\)      & \(3.15\times 10^{21}\)  & cm\(^{-3}\)        \\
\(\alpha\)         & 0.90                    & {\textendash} \\
\(C_{dir}^{hhe}\)  & \(8.75\times 10^{-32}\) & cm\(^6\)s\(^{-1}\) \\
\(C_{dir}^{hhe*}\) & 0.999902                & {\textendash} \\
\(p^*_{dir}\)      & \(1.47\times 10^{22}\)  & cm\(^{-3}\)        \\
\(\beta\)          & 1.55                    & {\textendash} \\
\(C_{pa}^{eeh}\)   & \(2.60\times 10^{-31}\) & cm\(^6\)s\(^{-1}\) \\
\(n^*_{pa}\)       & \(1.97\times 10^{19}\)  & cm\(^{-3}\)        \\
\(\gamma\)         & 0.79                    & {\textendash} \\
\(C_{pa}^{hhe}\)   & \(2.12\times 10^{-31}\) & cm\(^6\)s\(^{-1}\) \\
\(p^*_{pa}\)       & \(4.41\times 10^{19}\)  & cm\(^{-3}\)        \\
\(\delta\)         & 0.78                    & {\textendash}               
\end{tabular}
\caption{Parameters for Eq. \ref{eq:CvN}}
\label{tab:CvN}
\end{table}

For the temperature dependence at a carrier concentration of \oee, we construct our model based on Arrhenius activation for the direct AMR process,\cite{McAllister2015} while the \pa process is constructed by considering the Bose-Einstein occupation of phonons, with contributions from the low and high-energy peaks that we observe in Fig. 4c: 
\begin{equation}
\begin{split}
    C(T) & = C_{dir}^{eeh}e^{\frac{-E_a^{eeh}}{k_BT}} + C_{dir}^{hhe}e^{\frac{-E_a^{hhe}}{k_BT}} \\ 
    &\quad+\frac{C_{1,abs}^{eeh}}{e^{\frac{\hbar\omega_{\mathrm{low}}}{k_BT}}-1} + \frac{C_{2,abs}^{eeh}}{e^{\frac{\hbar\omega_{\mathrm{high}}}{k_BT}}-1} + C_{1,emit}^{eeh}\biggl(1+\frac{1}{e^{\frac{\hbar\omega_{\mathrm{low}}}{k_BT}}-1}\biggr) + C_{2,emit}^{eeh}\biggl(1+\frac{1}{e^{\frac{\hbar\omega_{\mathrm{high}}}{k_BT}}-1}\biggr) \\
    &\quad+ \frac{C_{1,abs}^{hhe}}{e^{\frac{\hbar\omega_{\mathrm{low}}}{k_BT}}-1} + \frac{C_{2,abs}^{hhe}}{e^{\frac{\hbar\omega_{\mathrm{high}}}{k_BT}}-1} + C_{1,emit}^{hhe}\biggl(1+\frac{1}{e^{\frac{\hbar\omega_{\mathrm{low}}}{k_BT}}-1}\biggr) + C_{2,emit}^{hhe}\biggl(1+\frac{1}{e^{\frac{\hbar\omega_{\mathrm{high}}}{k_BT}}-1}\biggr),
    \label{eq:CvT}
\end{split}
\end{equation}
where the corresponding parameters are given in Table \ref{tab:CvT}. Given the carrier concentration of \oee, it is not necessary to include any Coulomb enhancement parameters in this equation.


\begin{table}[h]
\centering
\begin{tabular}{ccc}
Parameter                        & Value                   & Units \\ \hline
\(C_{dir}^{eeh}\)                & \(9.46\times 10^{-32}\) & cm\(^6\)s\(^{-1}\) \\
\(E_a^{eeh}\)                    & 0.0028                  & eV \\
\(C_{dir}^{hhe}\)                & \(1.02\times 10^{-31}\) & cm\(^6\)s\(^{-1}\) \\
\(E_a^{hhe}\)                    & 0.25                    & eV \\
\(\hbar\omega_{\mathrm{low}}\)   & 0.018                   & eV  \\
\(\hbar\omega_{\mathrm{high}}\)  & 0.059                   & eV  \\
\(C_{1,abs}^{eeh}\)              & \(5.47\times 10^{-32}\) & cm\(^6\)s\(^{-1}\) \\
\(C_{2,abs}^{eeh}\)              & \(4.43\times 10^{-32}\) & cm\(^6\)s\(^{-1}\) \\
\(C_{1,emit}^{eeh}\)             & \(8.69\times 10^{-32}\) & cm\(^6\)s\(^{-1}\) \\
\(C_{2,emit}^{eeh}\)             & \(1.22\times 10^{-36}\) & cm\(^6\)s\(^{-1}\) \\
\(C_{1,abs}^{hhe}\)              & \(6.12\times 10^{-32}\) & cm\(^6\)s\(^{-1}\) \\
\(C_{2,abs}^{hhe}\)              & \(5.47\times 10^{-32}\) & cm\(^6\)s\(^{-1}\) \\
\(C_{1,emit}^{hhe}\)             & \(5.58\times 10^{-32}\) & cm\(^6\)s\(^{-1}\) \\
\(C_{2,emit}^{hhe}\)             & \(2.09\times 10^{-32}\) & cm\(^6\)s\(^{-1}\)
\end{tabular}
\caption{Parameters for Eq. \ref{eq:CvT}}
\label{tab:CvT}
\end{table}

\begin{center}
\textbf{Visualization of Contributing Phonon Modes to Phonon-assisted AMR}
\end{center}

\begin{figure}[h]
\includegraphics{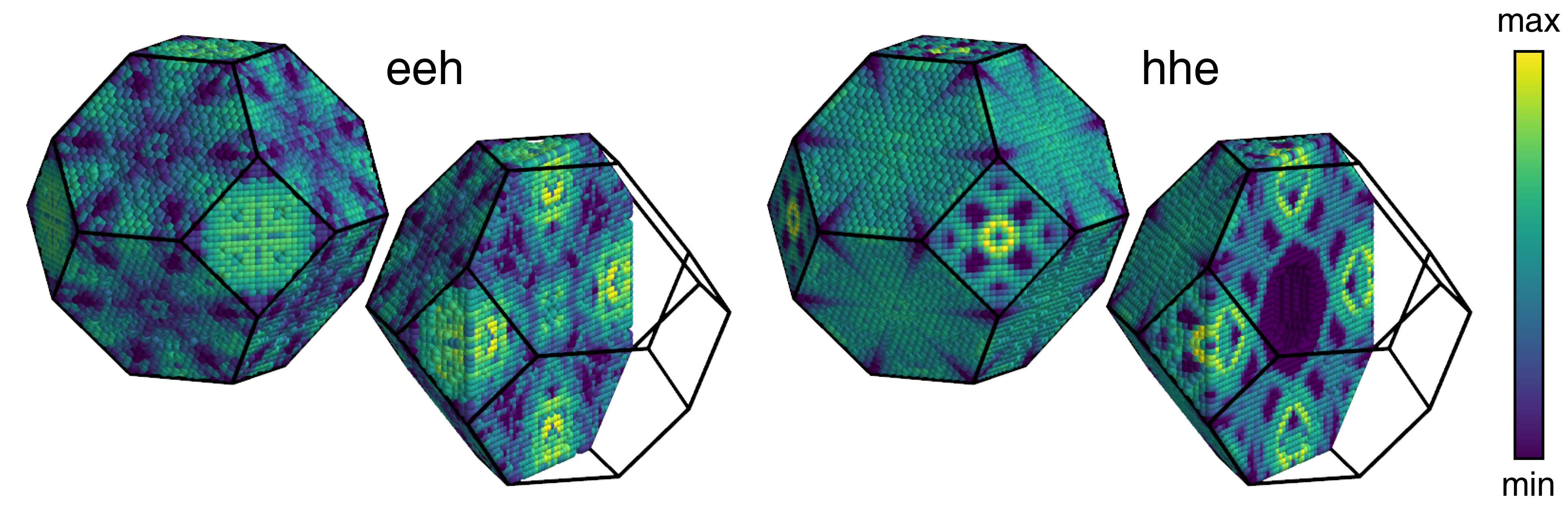}
\caption{The influence of different phonon wave vectors throughout the first BZ for the \(eeh\) and \(hhe\) processes.}
\label{fig:phonon_bz}
\end{figure}

In addition to the magnitude of the phonon wave vector (Fig. 4d), we can also visualize the distribution of contributing wave vectors in the first BZ as shown in Fig. \ref{fig:phonon_bz}, allowing for a more intuitive understanding of how different phonons contribute to the \pa \am process. For example, we clearly see two clusters around the \(X\) face and the \(\Gamma\) point of the BZ, which dominate for \(eeh\) \pa AMR. While for \(hhe\) \pa AMR, a broader {\textendash} though non-uniform {\textendash} distribution of short wavelength phonons from across the BZ surface participate, while long wavelength phonons are forbidden by momentum conservation.


\end{document}